\pdfoutput=1
\documentclass[a4paper,10pt]{article}

\usepackage{graphicx}
\usepackage{psfrag}
\usepackage{subfigure}
\usepackage[colorlinks=true
,urlcolor=blue
,anchorcolor=blue
,citecolor=blue
,filecolor=blue
,linkcolor=black
,menucolor=blue
,pagecolor=blue
,linktocpage=true
]{hyperref}

\usepackage[super,nonamebreak]{natbib}
\usepackage{amsmath}
\usepackage{amssymb}
\usepackage{color}
\usepackage{epic,eepic,rotating}

\title{\textbf {\begin{LARGE}Solving Hodgkin-Huxley equations using the compact difference scheme - tapering dendrite \end{LARGE}}}

\author{\begin{normalsize}%
  Asha Gopinathan\thanks{PI,Dendritic Simulator 
           project,Department of Neurology,Sree Chitra Tirunal Institute for 
           Medical Sciences and Technology,Trivandrum 695011,India.
           Mob: +919633568106, E-mail: dendron.15@gmail.com} 
  \& Joseph Mathew\thanks{Department of Aerospace Engineering, 
          Indian Institute of Science, Bangalore 560012, India}%
  \end{normalsize}}

\date{ }

\begin{document}

\maketitle

\title{}

\author{}

\newpage

\section*{\begin{large}\textbf{Summary}\end{large}}
Dendritic processing is now considered to be important in pre-processing of signals coming into a cell. Dendrites are involved in both propagation and backpropagation of signals$^{1}$. In a cylindrical dendrite, signals moving in either direction will be similar. However, if the dendrites taper, then this is not the case any more. The picture gets more complex if the ion channel distribution along the dendrite is also non-uniform. These equations have been solved using the Chebyshev pseudo-spectral method $^{2}$. Here we look at non-uniform dendritic voltage gated channels in both cylindrical and tapering dendrites.  For back-propagating signals, the signal is accentuated in the case of tapering dendrites. We assume a Hodgkin-Huxley formulation of ion channels and solve these equations with the compact finite-difference scheme. The scheme gives spectral-like spatial resolution while being easier to solve than spectral methods. We show that the scheme is able to reproduce the results obtained from spectral methods. The compact difference scheme is widely used to study turbulence in airflow, however it is being used for the first time in our laboratory to solve the equations involving transmission of signals in the brain.  
\section*{\begin {normalsize}\textbf{Introduction}\end{normalsize}}
Dendritic morphology plays an important role in determining both orthodromic and antidromic propagation $^{3,4,5,6,7,8,9}$. Additionally, the distribution of channels especially non uniform sodium channels in the dendrites influences the propagation of the signal. It is a combination of passive cable properties of dendritic membranes, sodium channel density and the diameter of the dendrite that influences the spike initiation in dendrites$^{10,11}$. Additionally,tapering optimises charge transfer from all dendritic synapses to the dendritic root$^{12}$. \\
\parindent = 1 in In this paper we use the compact difference scheme to solve the Hodgkin Huxley equations as they pertain to a tapering unbranched dendrite with a point soma. There are two types of taper under consideration : linear and exponential. The distribution of sodium and potassium channels follows an exponentially decaying function along the dendrite. The problem has been solved using the Chebyshev pseudo-spectral method$^{2}$. Assuming that there are just sodium, potassium and leak channels, these equations take the following form 
\begin{normalsize}
 \begin{equation}
 C_{m}\frac{\partial{V}}{\partial{t}} = \gamma_{0}(x)\frac{\partial^{2}V}{\partial{x}^2} + \gamma_{1}(x)\frac{\partial V}{\partial x}-I_{ion} + I_{in}(x,t)
\end{equation}
\begin{equation}
\gamma_{0}(x) = \frac{1}{2Ri }\frac{ r(x)}{\sqrt(1+r'^{2}(x))}
\end{equation}
\begin{equation}
 \gamma_{1}(x)= \frac{1}{Ri}\frac{ r'(x)}{\sqrt(1+r'^{2}(x))}
\end{equation}
where $r(x)$ is the radius of the dendrite, $r'(x) = dr/dx$, $C_{m}$ is the constant membrane capacitance,$R_{i}$ is the constant axial resistivity,$I_{ion}= I_{Na}+I_{K}+I_{L}$ and $I_{in}$ is the injected current.\\
\begin{equation}
 I_{Na}= g_{Na}(x)m^{3}h(V-E_{Na}), I_{K}= g_{K}(x)n^{4}(V-E_{K}), I_{L}= g_{L}(V-V_{L})
\end{equation} 
Equations for evaluating $g_{Na}$ and $g_{K}$ are given in reference $2$.\\
 The evolution equations for the potassium activation particle $n$ and sodium activation particle $m$ and inactivation particle $h$ are given by :
\begin{equation}
 \frac{dn}{dt} = \alpha _{n}(V)(1-n)-\beta_{n}(V)n
\end{equation}
\begin{equation}
 \frac{dm}{dt} = \alpha_{m}(V)(1-m)-\beta_{m}(V)m
\end{equation}
\begin{equation}
 \frac{dh}{dt} = \alpha_{h}(V)(1-h)-\beta_{h}(V)h
\end{equation}
$\alpha_{m}$, $\beta_{m}$, $\alpha_{h}$, $\beta_{h}$, $\alpha_{n}$, $\beta_{n}$ are evaluated from formulae given in reference $2$.\\
When current is injected at the point soma ( x = 0) only, $I_{in}$ is omitted from equation $1$ but appears in the boundary condition.Then in nondimensional form equation $1$ is:
\begin{equation}
\frac{ C_{m}}{\tau_{m}}\frac{\partial V}{\partial T} = \frac{\gamma_{0}(x)}{\lambda^{2}}\frac{\partial^{2} V}{\partial X^{2}} + \frac{\gamma_{1}(x)}{\lambda}\frac{\partial V}{\partial X}- I_{ion}
\end{equation} 
where $ T= t/\tau_{m}$, $ X= x/\lambda$, $\lambda = \lambda_{nontapered}(1+((2 \rho x)/d_{1}))^{1/2}$cm, $\rho$ is defined in equations $13, 15$, $d_{1}$ is the diameter at the nontapering end, $\tau_{m} = R_{m}C_{m}10^{3}$msec. 
\end{normalsize}
\section*{\begin{normalsize}\textbf{Spatial discretisation : Using compact finite difference schemes to solve the cable equation}\end{normalsize}}
Both the first and second derivative are approximated using the compact difference scheme. The equations used for  approximating the second derivative are given in reference $14$. Here we describe the equations used to approximate $V'$.
\begin{eqnarray}
\beta V'_{i-2} + \alpha V'_{i-1} + V'_{i} + \alpha V'_{i+1} + \beta V'_{i+2}& = & \frac{c (V_{i+3}-V_{i-3})}{6h}\nonumber \\ & &+\frac{b(V_{i+2}-+V_{i-2})}{4h} \nonumber \\
& & +\frac{ a( V_{i+1}-V_{i-1})}{h}, \\& &  (2\leq i\leq N -1 \nonumber )
\end{eqnarray} 
where $V'_{i}$ represents the finite difference approximation to the first derivative at node $i$ and $N$ is the maximum number of nodes in any given grid. 
 The relations between the coefficients $a$,$b$,$c$ and $\alpha$,$\beta$ are derived by matching the Taylor series coefficients of various orders. 
We take (ref.$13$,equation $2.1.7$)
\begin{displaymath}
\alpha = \frac{1}{3}, \beta = 0, a = \frac{14}{9}, b =\frac{ 1}{9}, c= 0
\end{displaymath}
When $\alpha = 1/3 $, the leading order truncation error becomes $ \frac{4}{7!} h^{6}d^7V/dx^7 $ making it sixth - order accurate.(ref.$13$ - Table 1) 
For boundaries the formula chosen is (ref. $13$, equation $4.1.1$):  
\begin{equation}
V''_{1}+ \alpha V''_{2} = \frac{ aV_{1}+bV_{2}+cV_{3}+dV_{4}}{h}
\end{equation} 
For third order accuracy, the coefficients are (ref.$13$, $ 4.1.3$) : 
\begin{eqnarray*} 
a = \frac{(11 + 2\alpha)}{6}, \hspace{2mm} b = \frac{ (6- \alpha )}{2} ,\hspace{2mm} c = \frac{(2\alpha -3)}{2}, \hspace{2mm} d =\frac{(2 -\alpha)}{6}
\end{eqnarray*}
The leading order truncation error ( on the r.h.s of equation $10$) is $(2(\alpha -3)/4!)h^{3}d^4V/dx^4)$. \\
Equations $ 10$, $11$ from reference $14$ and equations $9$ and $10$ applied at interior points results in a matrix problem $\textbf{A}V'' = \textbf{B}$ where A is tridiagonal and $V''$ can be obtained easily.
\section*{\begin{normalsize}\textbf{Time discretisation}\end{normalsize}}
 The values for $V''$ and $V'$ calculated from the compact-difference scheme were used to integrate the result in time using an explicit time stepping scheme - forward Euler. The scheme and its implementation is discussed in reference $14$.  Stability conditions requires the choice of the time step to be 
 \begin{equation}
\Delta T < \frac{\Delta X^{2}C_{m}\lambda^{2}}{\tau_{m}\gamma_{0}(X=0)}                                                                                                                                                                                                                                                                                                                                                                                                                                                                                                                                                   
  \end{equation} 
$\gamma_{0}(X=0)$ is maximum over the dendrite.$\Delta T$ varies as shown in ( ref.$14$, Table $1$).The numerical integration in time has been done with an explicit scheme. Since spatial derivatives are obtained with a compact scheme, which is an implicit formula that requires the solution of a linear system, implicit time-stepping is not possible. Implicit time-stepping is desirable in the case of stiff equations.  A work-around is to use a predictor-corrector scheme which uses an explicit step estimate from the predictor step in a corrector step which is also an explicit step.\\
Computations were performed on a Toshiba Satellite Pro laptop using Octave in a Linux(Ubuntu)environment. The data used for  simulations is given in (Table~\ref{tab:parameters}) and captions of (Fig.~\ref{fig:cabunbactap1}) and (Fig.~\ref{fig:cabunbactap2}).   
\section*{\begin{normalsize}\textbf{Configuration simulated}\end{normalsize}}
\begin{figure}[!ht]
\begin{center}
\subfigure[\textbf{Tapered dendrite construct with current injection at $i = 1$ and $ i = N$}]{
\includegraphics[width= 0.45\textwidth]{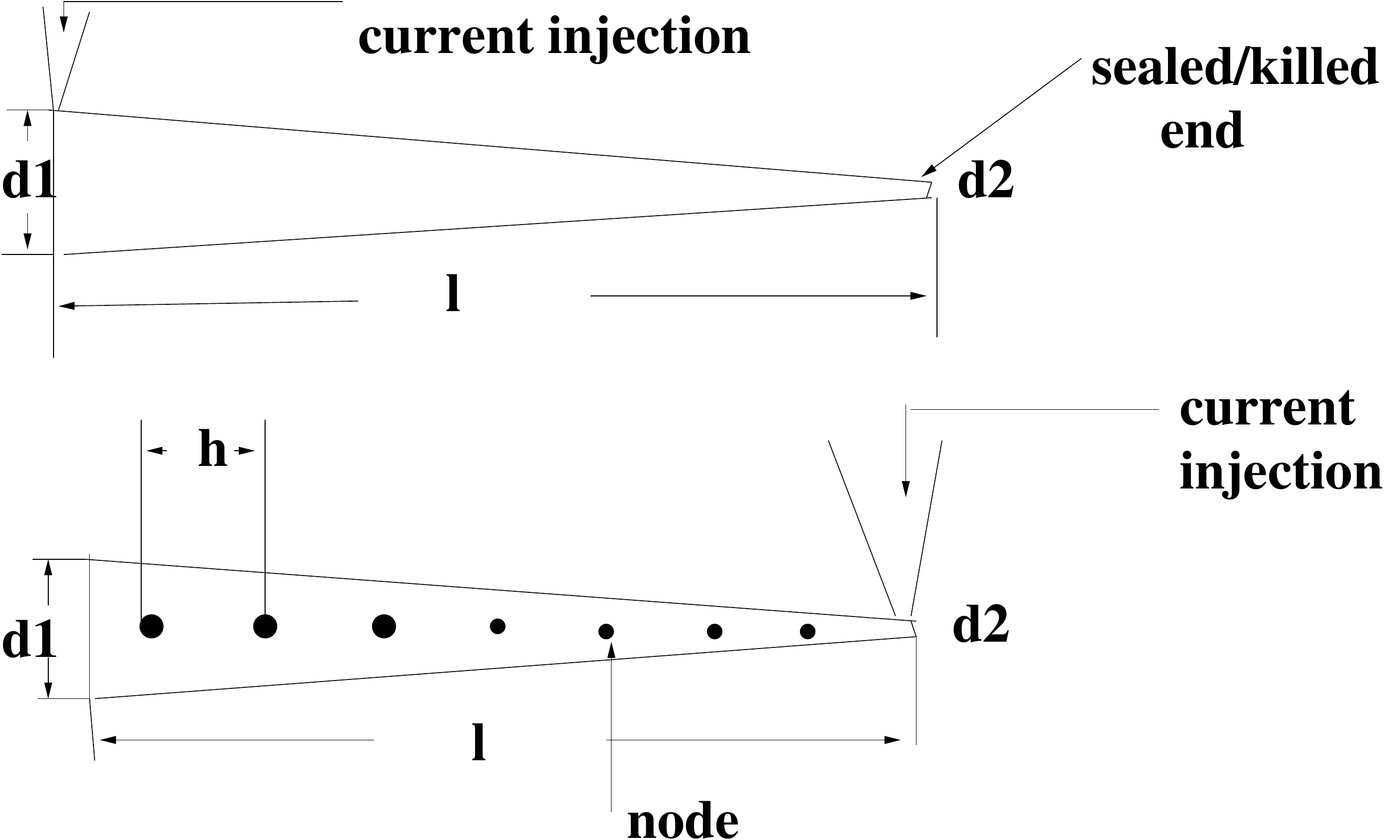}
}
\subfigure[\textbf{Equivalent circuit underlying the HH equations}]{
\includegraphics[width = 0.45\textwidth]{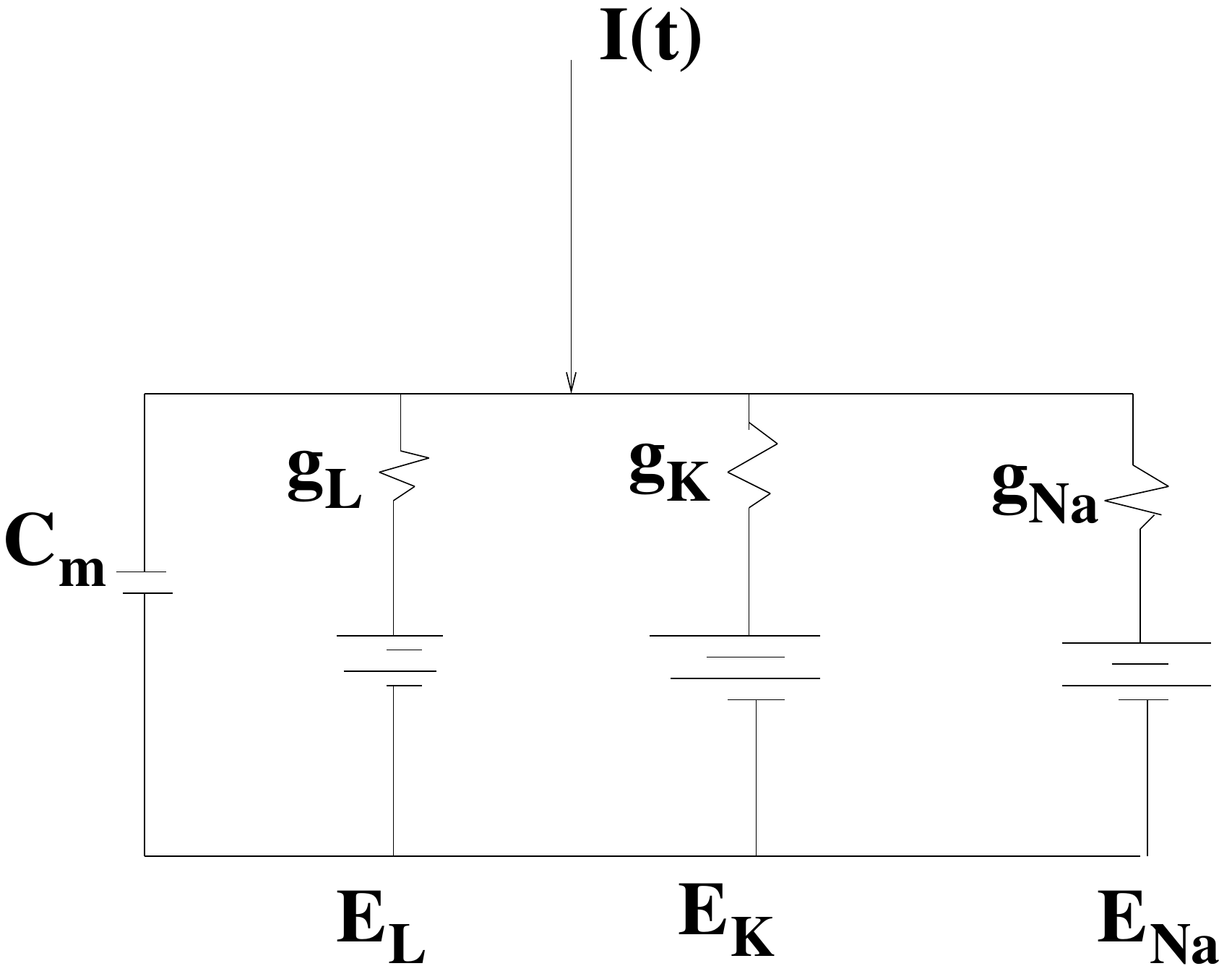}
}
\end{center}
\caption{\textbf{Soma tapering dendrite construct }}
\label{fig:somadendtapact}
\end{figure}
\begin{figure}[!ht]
\subfigure[\textbf{Cylindrical dendrite with stimulus on soma A }]{
\includegraphics[width = 0.45\textwidth]{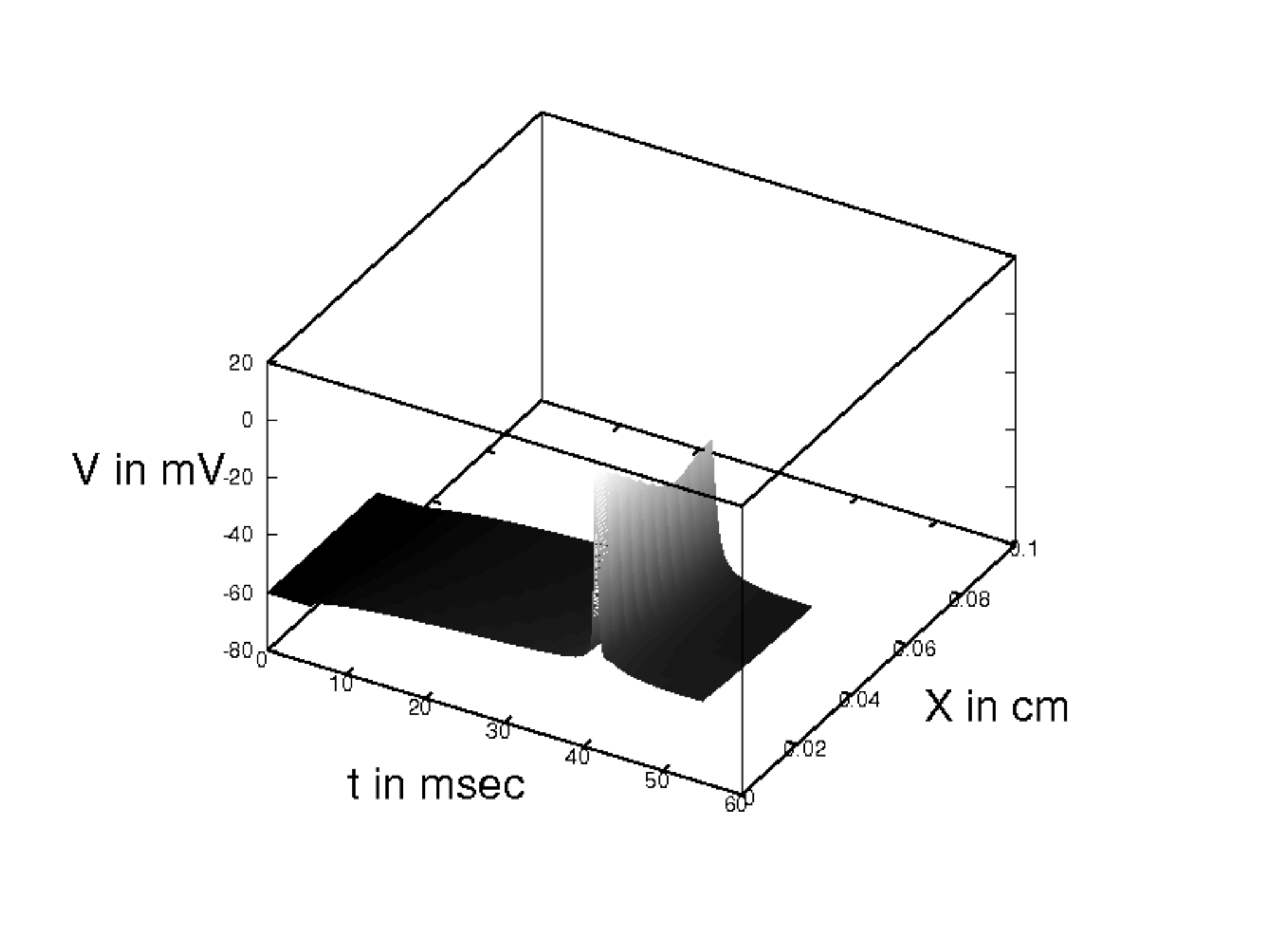}
}
\subfigure[\textbf{Cylindrical dendrite with stimulus on dendrite D}]{
\includegraphics[width = 0.45\textwidth]{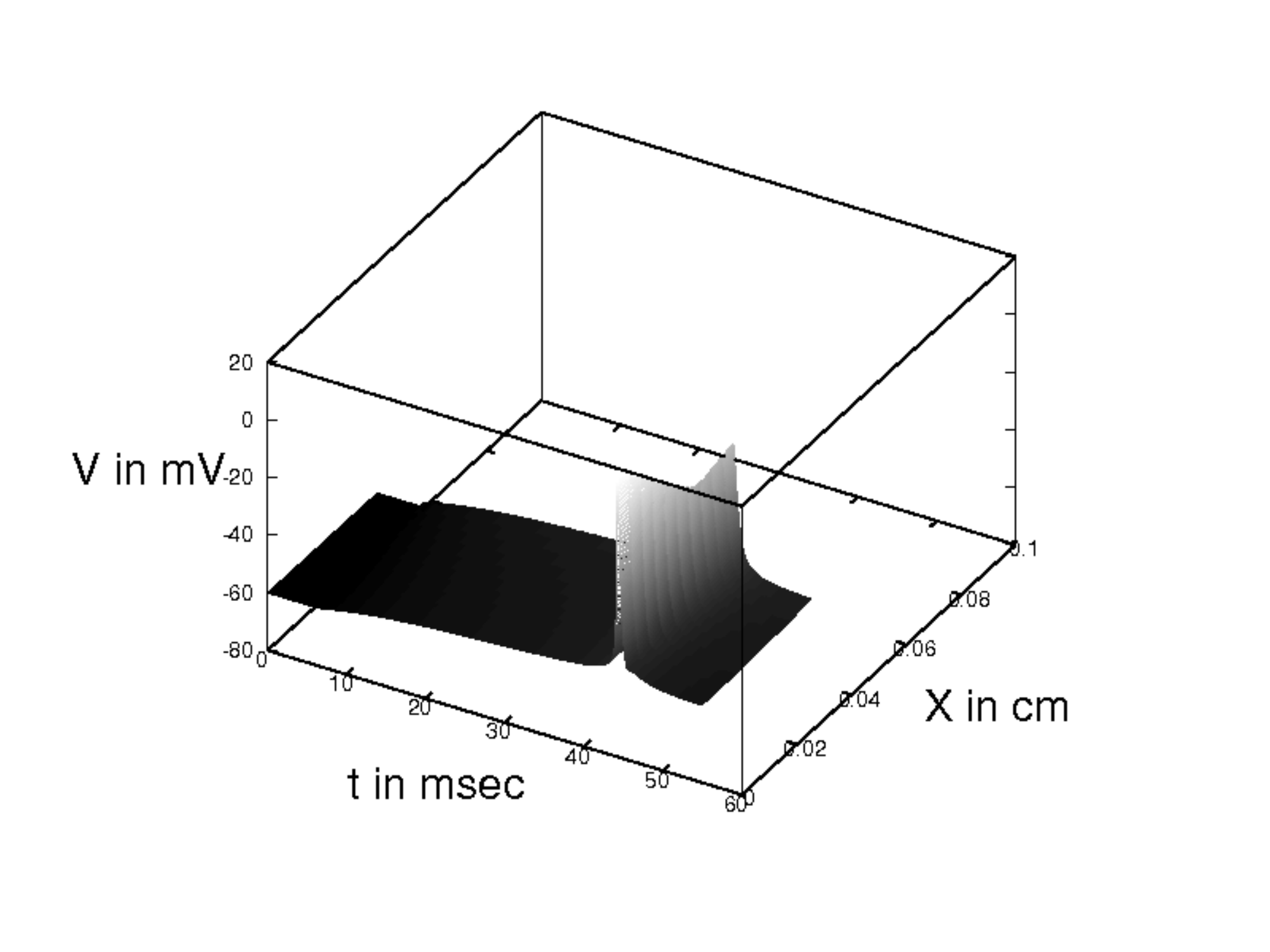}
}\\[3ex]
\subfigure[\textbf{Linear tapering with stimulus on soma B}]{
\includegraphics[width = 0.45\textwidth]{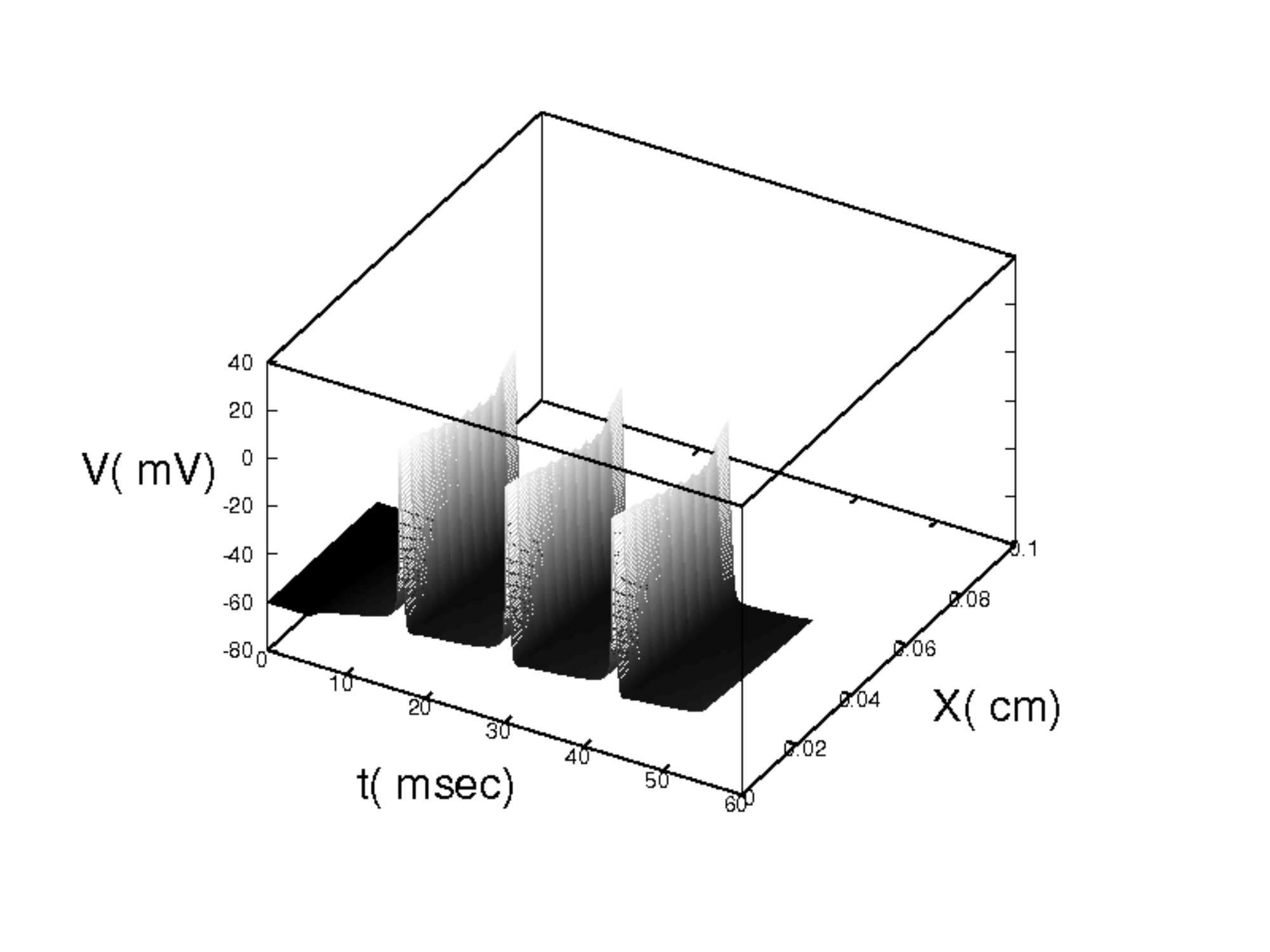}
}
\subfigure[\textbf{Linear tapering with stimulus on dendrite E}]{
\includegraphics[width = 0.45\textwidth]{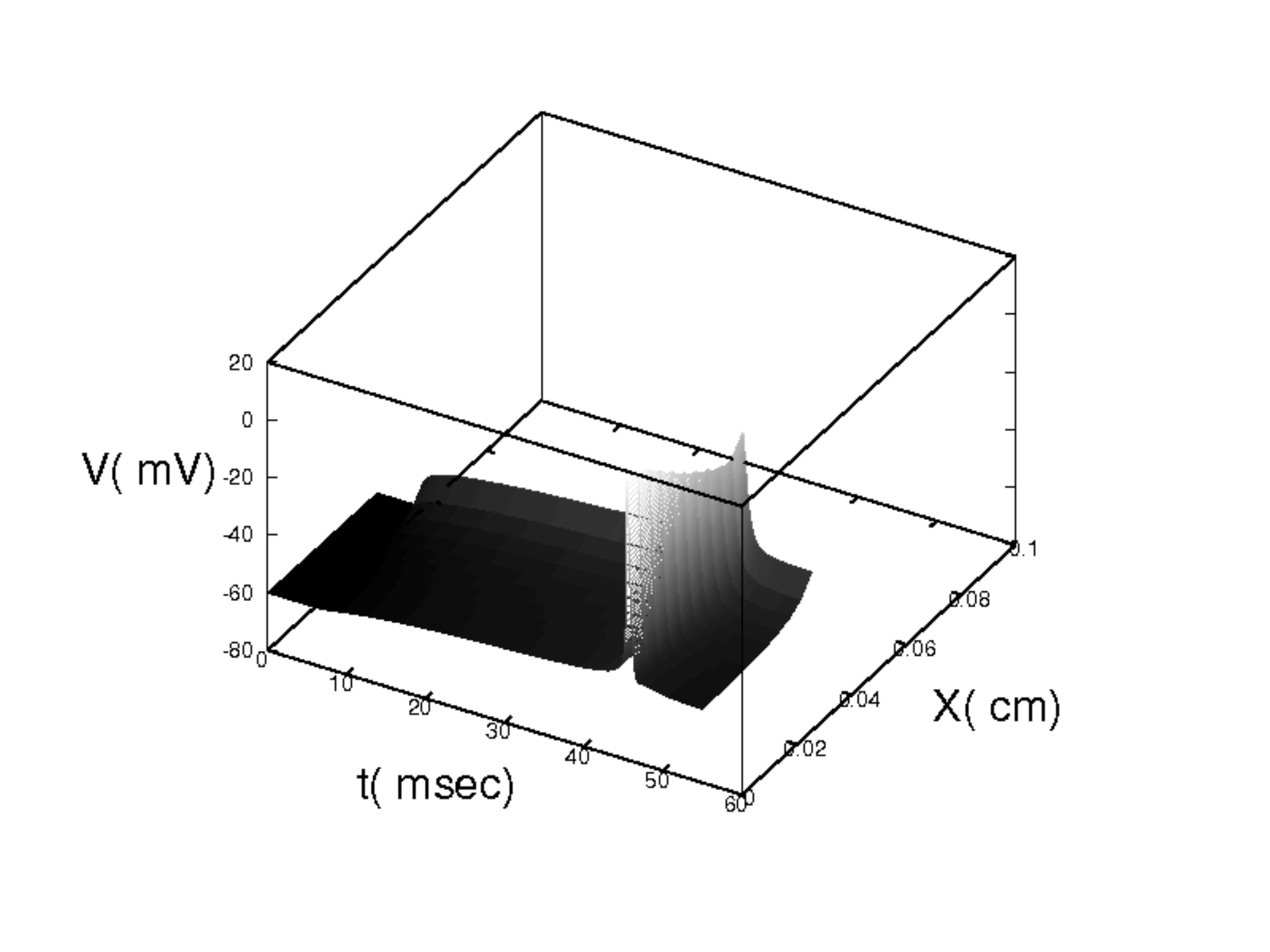} 
}\\[3ex]
\subfigure[\textbf{Exponential tapering with stimulus on soma C }]{
\includegraphics[width = 0.45\textwidth]{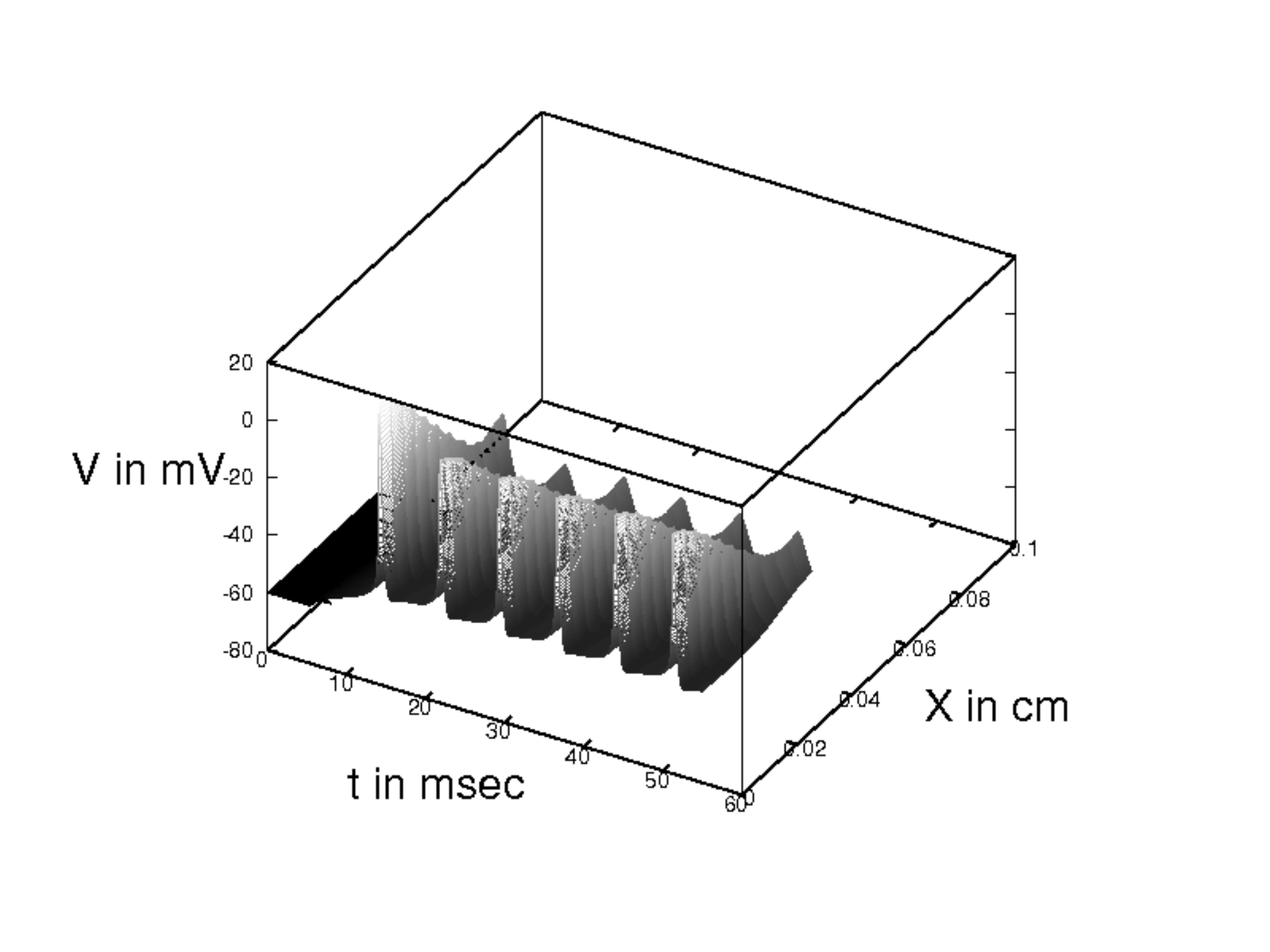}
}
\subfigure[\textbf{Exponential tapering with stimulus on dendrite F}]{
\includegraphics[width = 0.45\textwidth]{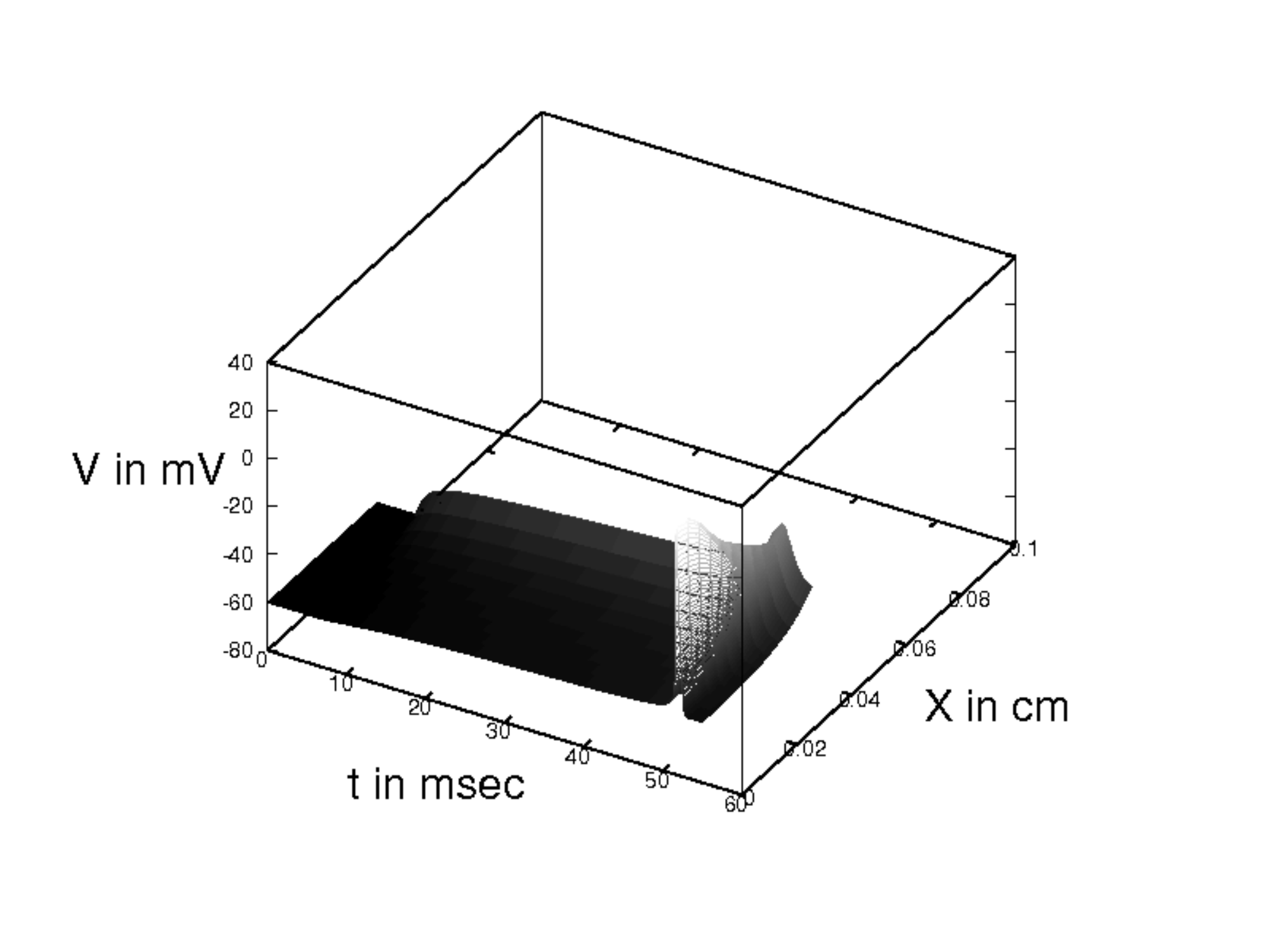} 
} 
\caption{\textbf{Simulation with differing geometery and location of stimulus. Exponentially distributed Na and K channels with $\lambda_{Na} = \lambda_{K} = -0.0075\mu$m$^{-1}$, and $g_{Na0} = 50mS/cm^{2}$, $ g_{K0} = 12.5 mS/cm^{2}$, $g_{L} = 0.1 mS/cm^{2}$ was constant throughout the soma-dendrite length. Stimulus intensity at the soma ($0 \mu$m) is $300 \mu A/cm^{2}$ in all cases (A-C) while for stimulation at the end of the dendrite ($ 400 \mu$m) the following intensities were used : $ 338 \mu A/cm^{2}$ with cylindric geometry (D),and $ 3380 \mu A/cm^{2}$ and $ 4657 \mu A/cm^{2}$ with linear (E) and exponential tapering (F) respectively. The duration of the stimulus was $50$ms in all cases,starting at $5$ms.}}  
\label{fig:cabunbactap1}
\end{figure}
\begin{figure}[!ht]
\subfigure[\textbf{Cylindrical dendrite A }]{
\includegraphics[width = 0.45\textwidth]{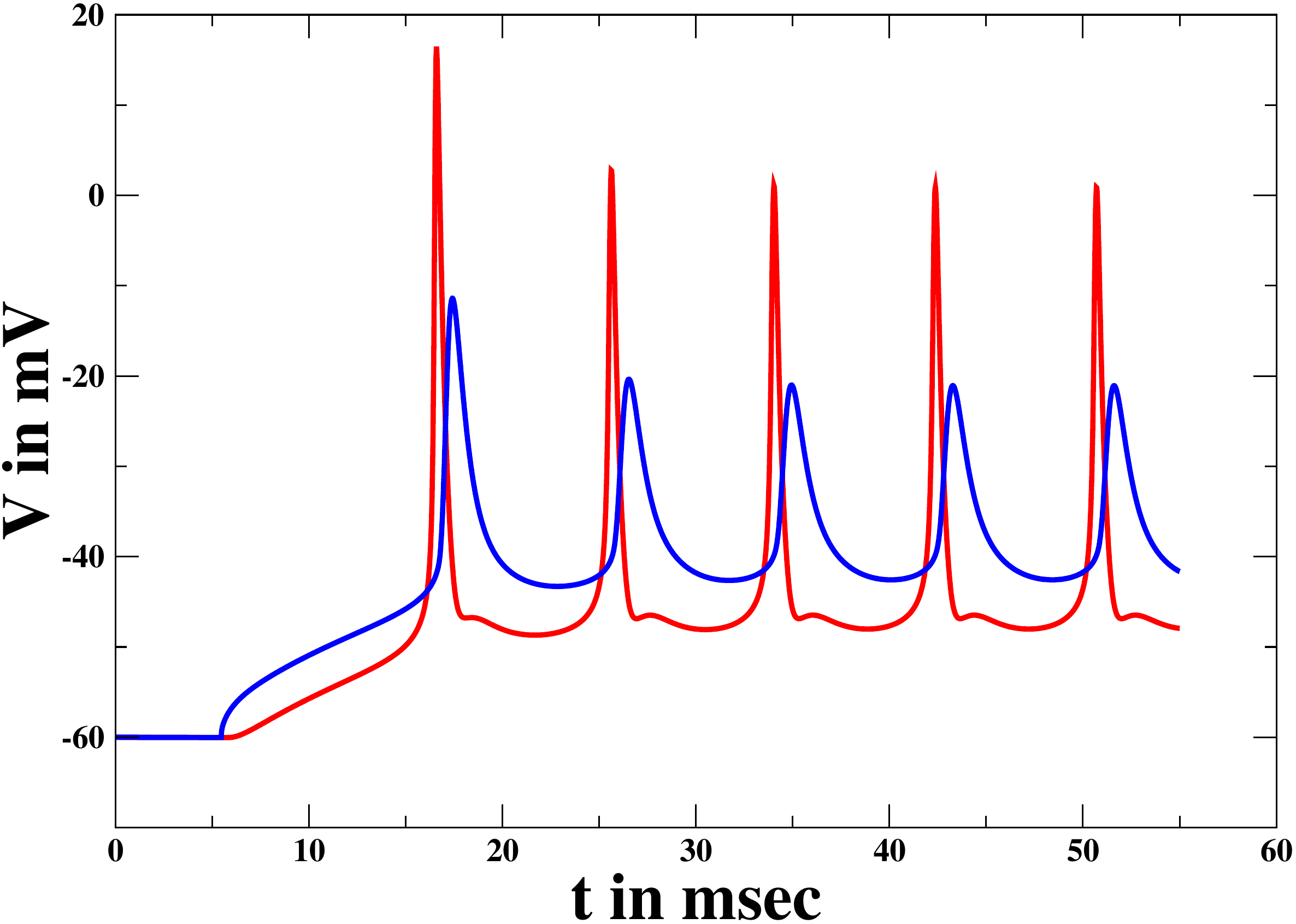}
}
\subfigure[\textbf{Linear tapering dendrite B}]{
\includegraphics[width = 0.45\textwidth]{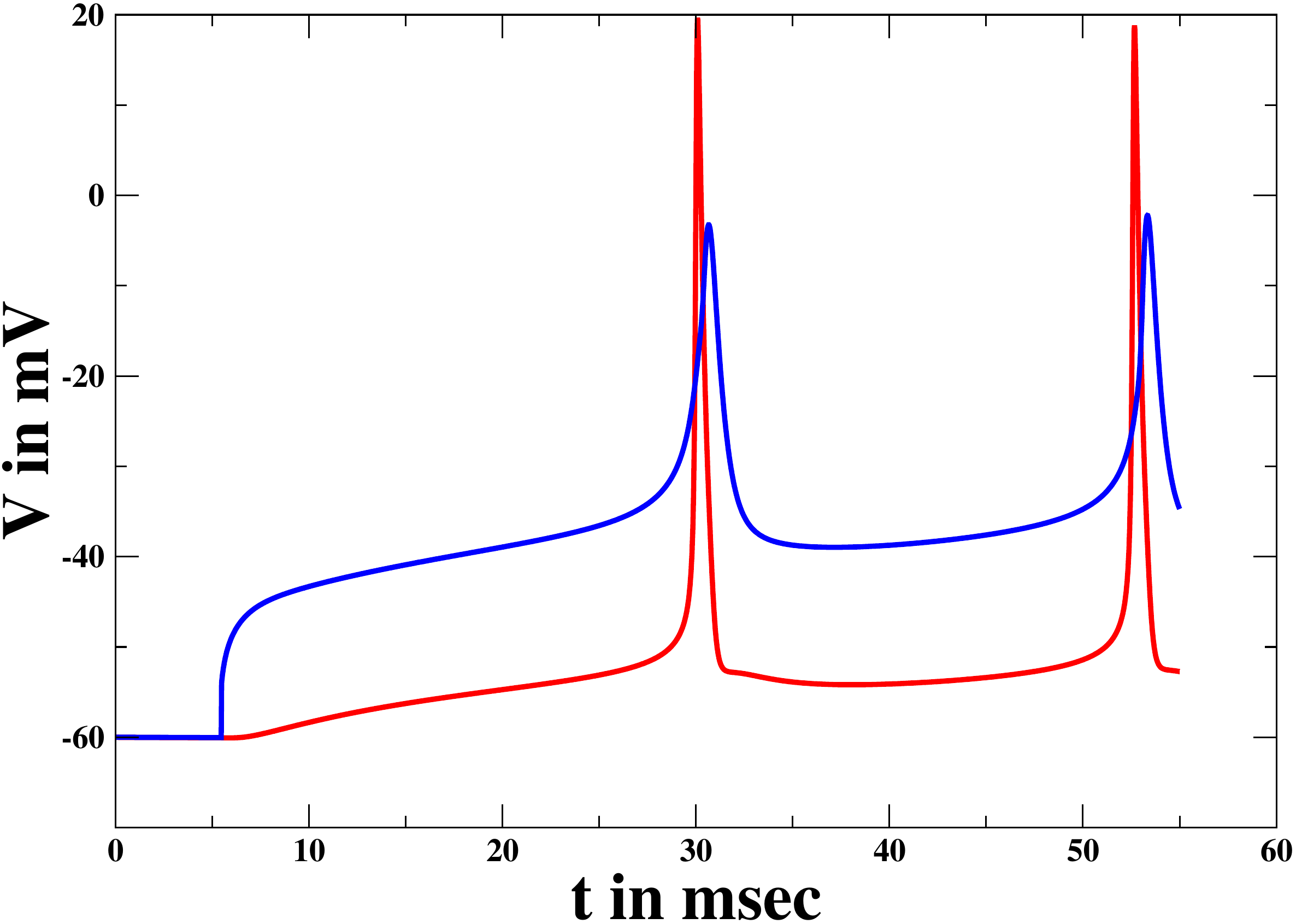}
}\\[3ex]
\subfigure[\textbf{Exponential tapering dendrite C}]{
\includegraphics[width = 0.45\textwidth]{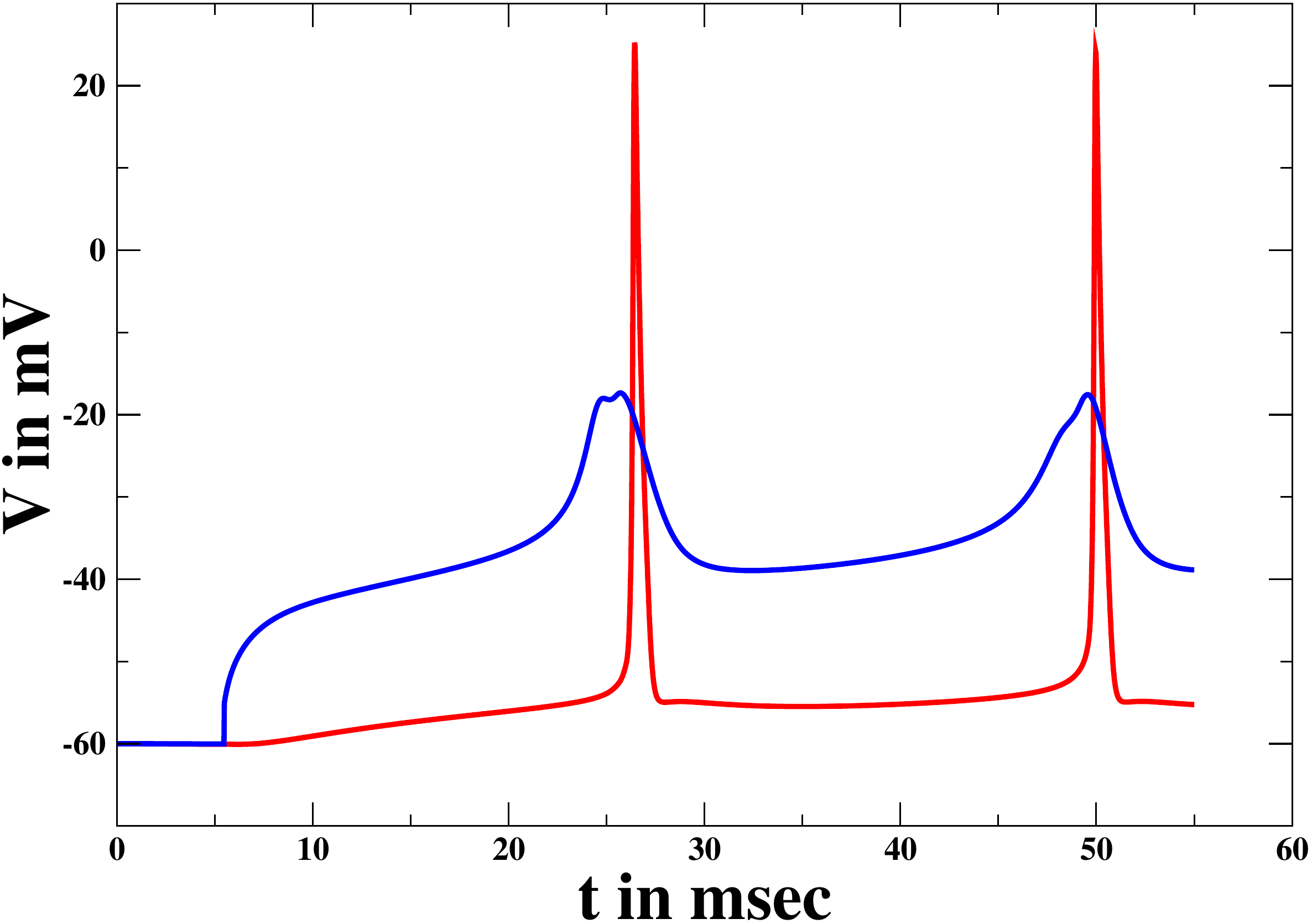}
}
\subfigure[\textbf{Exponential tapering dendrite with near threshold stimulus D}]{
\includegraphics[width = 0.45\textwidth]{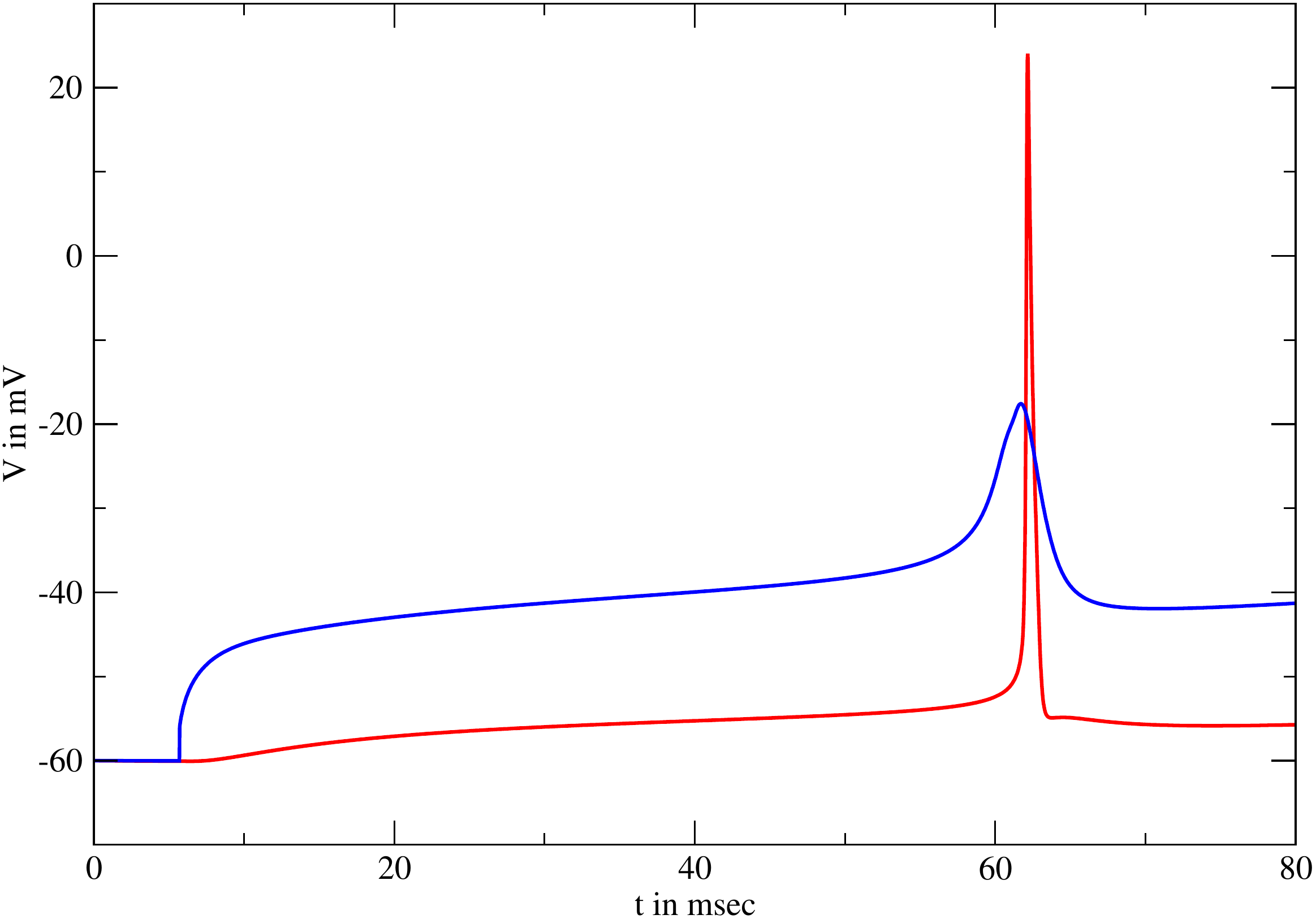} 
}\\
\caption{\textbf{Effects of dendritic geometry on the propagation and back-propagation of action potentials along the dendrite. In all cases, a dendritic stimulus of $50$ms duration, starting at $5$ms was applied. The amplitude changed in each case. ( A ) Cylindric geometry: a stimulus of $ 750 \mu A/cm^{2}$ . (B) Linear tapering of dendrite : stimulus intensity is $ 3750 \mu A/cm^{2}$. (C) Exponential tapering of dendrite : stimulus intensity is $ 5250 \mu A/cm^{2}$. (D) Exponential tapering : a near threshold stimulus $ 4393 \mu A/cm^{2}$. Note the different time scale on this record.{ soma, $ i = 1$,{\color{red} \textemdash} ;{end of dendrite,$ i = N$,{\color{blue} \textemdash }}}}}
\label{fig:cabunbactap2}  
\end{figure}

The two sets of problems attempted here are as follows : \\
In the first spatio temporal evolution of $V$ was examined in cylindrical, linearly and exponentially tapering dendrites. The linear taper of the dendrite is determined by 
\begin{equation}
 r(x) = \rho x + r(0)
\end{equation} 
\begin{equation}
 \rho = (r(L) -r(0))/L
\end{equation} 
where $ r(x)$ is the radius at any given point, $r(0)$ is the radius at $ x = 0$ and $r(L)$ is the radius of the dendrite at $ x = L$ and $ l$ is the length of the dendrite. Values for $r(0)$,$r(L)$ and $l$ are given in ( Table 1).\\ 
In the second case exponential taper is determined by:
\begin{equation}
 r(x) = r(0)exp(-\rho x)
\end{equation} 
\begin{equation}
 \rho = ln[r(0)/r(L)]/L
\end{equation} 
Initial and boundary conditions are the same as given in reference $14$.The current injection was initially on the  soma. This was followed by  injections on the dendrite. This is illustrated in (Fig.~\ref{fig:somadendtapact}).In the second case, the effect of dendritic geometry on propagation and back propagation of action potential was examined. Here too cylindrical, linearly and exponentially tapering dendrites were examined. The current was injected at the dendritic tip in all cases. \\
Results are shown in (Fig.~\ref{fig:cabunbactap1})and (Fig.~\ref{fig:cabunbactap2}). It can be seen in (Fig.~\ref{fig:cabunbactap1})that stimulus on the soma or tip of dendrite in a cylindrical dendrite leads to the cell firing once for the given parameters. However in the case of linear and exponential tapering, the cells fires multiple times when current is injected at the soma but only once when it is injected at the tip of the dendrite despite increasing the stimulus intensity by 16 ( paper uses 10 times) and 14 times for the linear and exponential tapering respectively. In (Fig.~\ref{fig:cabunbactap1})b,distribution of ion channels $\lambda_{Na}$,$\lambda_{K} $ used is $ -0.0025 \times 10^{4} $cm $^{-1}$ and in (Fig.~\ref{fig:cabunbactap1})c, $ \lambda_{Na} $,$ \lambda_{K} $ used is $-0.010 \times 10^{4}$cm$^{-1}$. These values which are slightly different from those used in reference $2 $ yield the same number of firing as shown in the reference. In (Fig.~\ref{fig:cabunbactap2}), the stimulus intensity is greater than that used in the earlier figure and elicits a train of action potentials which propagate to the soma and back-propagate to the dendritic stimulus site ( blue ).In (Fig.~\ref{fig:cabunbactap2})a, $\lambda_{Na}$, $\lambda_{K}$ used is $ -0.010 \times 10{^4}$ cm$^{-1}$ which is slightly higher than that used in reference $2$. This yields exactly the same number of firing as shown in (Fig.~\ref{fig:cabunbactap2})a. Here too tapering reduces the firing seen when stimulus is applied at the dendrite. It has also been observed that a much higher stimulus intensity than that reported in the paper has been used by us to get the desired result. 
\clearpage
\clearpage
\section*{Tables}
\begin{table}[h!]
\caption{\bf{Parameters of dendrite used in simulation}}
\centering
\begin{tabular}{c rr}
\hline 
Parameter& \multicolumn{2}{c}{Values} \\
\hline
length & $400 \mu$\\
diameter $d1$ & $3.7 \mu$ \\
diameter $d2$ & $0.3 \mu $ \\
$R_{m}$ & $10^{-3} \Omega.cm^{2} $  \\
$R_{i}$ & $330  \Omega$.cm \\
$C_{m}$ & $1^-6 farad/cm^{2}$\\
$\tau$ & $1^{-5}$ sec \\
$V_{Na}$& $55 $ mV \\
$V_{K}$ & $ -95$ mV \\
$V_{L}$&$-60$ mV\\
\hline
\end{tabular} 
\label{tab:parameters}
\end{table}
\begin{table}[h!]
\caption{Resolving Efficiency $\epsilon$ of the first derivative schemes,(ref.${13}$, Table $4$)}
\centering
\begin{tabular}{c c c c }
\hline\hline
Scheme & $\epsilon = 0.1$ & $\epsilon = 0.01$ & $\epsilon = 0.001$  \\[0.5ex]
\hline
Fourth order central & 0.44& 0.23 & 0.13\\
Fourth order compact  & 0.59 & 0.35 & 0.20 \\
Sixth order tridiagonal & 0.70 & 0.50 & 0.35 \\
\hline 
\end{tabular}
\label{tab:resolving}
\end{table}
\begin{table}[h!]
\caption{Injected current in (Fig.~\ref{fig:cabunbactap1}) and (Fig.~\ref{fig:cabunbactap2})}
\centering
\begin{tabular}{c c c}
\hline\hline
Figure & $I \mu A/cm{^2}$ ( paper) & $ I \mu A/cm{^2}$(actual) \\[0.5ex]
\hline
2 e  & 3380 & 5659  \\
3b  & 3750 & 6720  \\
3c & 5250 & 5482 \\
3d & 4393 & 4810 \\
\hline 
\end{tabular}
\label{tab:injval}
\end{table}
\section*{\begin{normalsize}\textbf{Concluding Remarks}\end{normalsize}} In this paper we have used the compact finite difference scheme to solve the Hodgkin Huxley equations for a tapering dendrite. It has been shown that the scheme is robust and can reproduce the results seen in reference $2 $ in (Fig.~\ref{fig:cabunbactap1}) and (Fig.~\ref{fig:cabunbactap2}). Convergence as defined in reference $14$ has been tested in (Fig.~\ref{fig:cabunbactap2}) ( a, b,c ) and it is seen that the solution is convergent. In all cases of injection of current from the dendritic end in tapering dendrites, the current used had to be changed slightly from that reported in reference $2 $. The values used by us are reported in (Table~\ref{tab:injval}). In (Fig.~\ref{fig:cabunbactap2}). the plots in red are those of the action potentials generated at the soma and that in blue of those generated at the dendrite. The dendritic action potential occurs after the somatic action potential in both cylindrical and linearly tapering dendrites. In the exponentially tapering dendrite (Fig.~\ref{fig:cabunbactap2}), it is seen that the first dendritic action potential occurs before the somatic action potential. The second one occurs closer to the somatic action potential. This indicates variation in travelling speed of excitation from dendrite to soma and back propagation. \\ We have provided here an alternate to the Chebyshev pseudo- spectral method used in reference $2$. This is an easier method to implement and can give spectral like resolution depending on the scheme chosen. As shown in Table~\ref{tab:resolving}, for the first derivative, at $\epsilon = 0.001$, the resolving efficiency of sixth order and fourth order compact difference is higher than that of a fourth order central difference scheme. By selecting different coefficients, this efficiency can be brought closer to exact differentiation. The resolving efficiency for second derivative schemes is shown in Table 3 (ref. $14$). Here too it can be seen that the at $\epsilon = 0.001$, the resolving efficiency of the sixth order tridiagonal and fourth order compact scheme is higher than that of the central difference scheme. \\As discussed in reference $14$, implicit methods for time stepping cannot be utilised. However, a corrector - predictor method can be used to circumvent the issues involving this.\\
As discussed earlier, certain values of $\lambda_{Na}$ and $\lambda_{K}$ have had to be altered along with the $I$ values. In reference $14$ it can be seen that both compact and fourth order central schemes are affected similarly by these changes. Thus it is not scheme specific. \\This paper along with reference $14$ thus shows that the compact scheme can be used as an alternate method to solve the HH equations in both cylindrical and tapering dendrites. By picking different coefficients, the spatial resolution at any given error can be improved. It is our understanding that this is the first time it has been used to solve these equations.  
\section*{\begin{normalsize}\textbf{References}\end{normalsize}}
1. Rinzel,J. Distinctive roles for dendritic neuronal computation.\textit{SIAM News}.\textbf{40:2}, 2007.\\
2. Toth,T.I \& Crunelli,V. Effects of tapering geometry and inhomogeneous ion channel distribution in a neuron model. \textit Neuroscience. \textbf{84:4}, 1223-1232, 1998.\\
3. Vetter,P, Roth,A \& Rall,W. Propagation of action potentials in dendrites depends on dendritic morphology. \textit {J.Neurophysiol}. \textbf {85}, 926-937, 2001. \\
4. Mainen,Z.F \& Sejnowski,T.J. Influence of dendritic structure on firing pattern in model neocortical neurons. \textit {Nature}. \textbf{382}, 363-366, 1996.\\
5. Hausser,M,Stuart,G,Racca C,\& Sakmann,B. Axonal initiation and active dendritic propagation of action potentials in substantia nigra neurons. \textit {Neuron}. \textbf{15}, 637-647, 1995.\\
6. Stuart,G \& Hausser,M. Initiation and spread of sodium action potentials in cerebellar Purkinje cells. \textit {Neuron}. \textbf{13}, 703-712, 1994.\\
7. Stuart,G \& Sakmann,B. Active propagation of somatic action potentials into neocortical pyramidal cell dendrites. \textit {Nature}. \textbf{367}, 69-72, 1994.\\
8. Stuart,G , Schiller, J \& Sakmann, B. Action potential initiation and propagation in rat neocortical pyramidal neurons. \textit {J Physiol(Lond)}. \textbf{505}, 617-632, 1997a. \\
9. Stuart,G , Spruston,N, Sakmann,B \& Hausser,M. Action potential initiation and backpropagation in neurons of the mammalian CNS. \textit {Trends in Neurosci.} \textbf{20}, 125-131, 1997b.\\
10. Rapp,M, Yarom, Y \& Segev,I. Modeling back propagating action potential in weakly excitable dendrites of neocortical pyramidal cells. \textit Proc.Natl.Acad.Sci U.S.A. \textbf{93}, 11985-11990, 1996.\\
11. Williams,S.R \& Stuart, G.J. Action potential back propagation and Somato-dendritic distribution of ion channels in thalamocortical neurons. \textit The Journal of Neuroscience. \textbf{20:4}, 1307-1317, 2000. \\
12.Cuntz,H, Borst, A  and Segev,I. Optimization principles of dendritic structure. \textit Theoretical Biology and Medical Modelling. \textbf{21:4}, 2007.\\
13.Lele,S.K. Compact finite difference schemes with spectral-like resolution. \textit J.Comp Phy. \textbf{103}, 16-42, 1992. \\
14. Gopinathan,A \& Mathew,J.Solving Hodgkin-Huxley equations using the compact difference scheme - somadendrite. (http://arxiv.org/abs/1308.1522).\\  
\textbf{Supplementary Information} is available in the online version of the paper.\\
\textbf{Acknowledgements} AG would like to acknowledge the support provided by A.K Gupta( currently NIMHANS, Bangalore)and M.D Nair of Sree Chitra Tirunal Institute of Medical Sciences and Technology, Tiruvananthapuram.  AG would also like to acknowledge the support of  V.Nanjundiah in making arrangements to work at the Indian Institute of Science. AG thanks Elizabeth Jacob for support provided at NIIST, Trivandrum during the writing of this paper. AG thanks Maya Ramachandran and Venugopalan for acquiring necessary references from the library of the National University of Singapore. Thanks are due to Ganesh of SPACE, Tiruvananthapuram for help with opensource software.AG also acknowledges the ready help provided by the octave- users group in solving any problems that have arisen with the Octave code. AG is supported by the DST- WOS-A grant which covered the costs of this project. \\
\textbf{Author Contributions} JM suggested the use of the compact difference scheme as an alternative to the spectral scheme and helped in smoothing out troubles during its implementation. AG ran the simulations, wrote the code and wrote the paper. Both authors interpreted the results and edited the papers. \\
\textbf{Author Information} Reprints and permission information is available at www.nature.com/reprints. The authors declare no competing financial interests. Readers are welcome to comment on the online version of the paper. Correspondence and requests for materials should be addressed to AG(dendron.15@gmail.com)\\
\end{document}